\documentstyle[pre,aps,epsf,multicol]{revtex}
\begin{document}
\draft

\hfuzz=100pt
\hbadness=10000


\title{Asymptotic step profiles from a nonlinear growth equation
for vicinal surfaces}

\author{Jouni Kallunki and Joachim Krug}
\address{Fachbereich Physik, Universit\"at GH Essen, 45117 Essen, Germany}
\date{\today}
\maketitle

\begin{abstract}
We study a recently proposed
nonlinear evolution equation describing the collective
step meander on a vicinal surface subject to the Bales-Zangwill
growth instability [O. Pierre-Louis {\it et al.}, Phys. Rev. Lett.
{\bf 80}, 4221 (1998)].
A careful numerical analysis shows that the
dynamically selected step profile consists of sloped segments, given
by an inverse error function and steepening
as $\sqrt{t}$, which are matched
to pieces of a stationary (time-independent) solution
describing the maxima and minima.
The effect of
smoothening by step edge diffusion is included heuristically,
and a one-parameter family of evolution equations is introduced
which contains relaxation by step edge diffusion and by
attachment-detachment as special cases.
The question of the persistence of an initially imposed
meander wavelength is investigated in relation to recent experiments.
\end{abstract}



\begin{multicols}{2}
\narrowtext

\section{Introduction}

Ten years ago, Bales and Zangwill \cite{Bales90}
predicted that a growing
vicinal surface should undergo a step meandering instability
when kinetic step edge barriers suppress the attachment of atoms
to descending steps \cite{Schwoebel66}. The instability has meanwhile
been observed in experiments \cite{Exps,Maroutian99}
and Monte Carlo simulations
\cite{Rost96}, and a number of theoretical studies have been devoted to the
nonlinear evolution of the surface both in the presence
\cite{Pierre-Louis98a} and
absence \cite{Rost96,Pierre-Louis98b} of desorption
\cite{Politi00}.

Since linear stability analysis shows the
in-phase mode of the collective step meander to be the most unstable
\cite{Pimpinelli94}, the two-dimensional surface morphology can
be represented by a one-dimensional function $\zeta(x,t)$ describing
the displacement of the common step profile from the flat straight
reference configuration $\zeta = 0$, with the $x$-axis oriented
along the step \cite{Note:Coherence}. For the case of
infinite step edge barriers, attachment-detachment kinetics
and no desorption, the nonlinear evolution equation
\begin{equation}
\label{st_eq}
\zeta_t = -\left\{ \frac{\alpha \zeta_x}{1+\zeta_x^2}  +
\frac{\beta}{1+\zeta_x^2}
\left[ \frac{\zeta_{xx}}{(1+\zeta_x^2)^{3/2}}
\right]_x \right\}_x
\end{equation}
was proposed in Ref.\cite{Pierre-Louis98b}
(subscripts denote derivatives).
It can be derived
from the Burton-Cabrera-Frank (BCF) theory of growth on vicinal surfaces
\cite{BCF51}
using a singular multiscale expansion
\cite{Pierre-Louis98b,Gillet00} in $\epsilon^{1/2}$,
where $\epsilon = \Omega F \ell^2/D$ is the P\'eclet number. Here
$F$ is the deposition flux,
$D$ the in-plane surface diffusion coefficient,
$\ell$ the nominal step spacing, and
$\Omega$ the atomic area. The coefficients in Eq.(\ref{st_eq})
are given by $\alpha = \Omega F \ell^2/2$ and
$\beta = \Omega^2 D \ell \gamma c_{\rm eq}/k_B T$, with
$\gamma$ and $c_{\rm eq}$ referring to the step stiffness and the
equilibrium adatom density, respectively.

According to (\ref{st_eq}), the straight step is linearly unstable
against perturbations with wavelengths larger than $\lambda_c =
2 \pi \sqrt{\beta/\alpha}$, with a fastest growing wavelength
$\lambda_u = \sqrt{2} \lambda_c$. To explore the
nonlinear regime, in Ref.\cite{Pierre-Louis98b}
a numerical integration of Eq.(\ref{st_eq}) was carried out which
showed an increase of the meander amplitude as $\sqrt{t}$ at
fixed wavelength $\lambda_u$, as well as the formation of spike
singularities at maxima and minima of $\zeta$. The latter is surprising
because the second term on the right hand side of (\ref{st_eq})
would be expected to suppress such rapid variations of the step curvature.

Here we revisit the problem using a more accurate numerical algorithm
\cite{Note:Gillet1}.
We demonstrate that the step profile remains smooth near maxima and
minima, where it approaches asymptotically a {\it stationary}
(time-independent) solution of (\ref{st_eq}), while the sides of the
profile follow a separable solution
with an amplitude of order $\sqrt{t}$.
The matching of the two solutions occurs near the point of maximum slope.
We further show heuristically how the effect of step edge diffusion
can be included in the theory, and introduce
a generalized evolution equation which contains edge diffusion and
attachment-detachment kinetics as special cases.
Finally, we address the question to what extent an initially imposed
meander wavelength different from $\lambda_u$ is preserved under the time
evolution. This is relevant in view of the recent
experiments of Maroutian {\it et al.} \cite{Maroutian99}.

\section{Shape selection}

Before presenting the numerical results we recapitulate the two
classes of analytic solutions to (\ref{st_eq}) which were
found in \cite{Pierre-Louis98b}.
{\it Stationary} solutions are obtained by setting the mass
current along the step (the quantity inside to curly brackets
on the right hand side of (\ref{st_eq})) to zero. In terms
of $m(x) = \zeta_x/\sqrt{1 + \zeta_x^2}$ the
stationarity condition reduces to Newton's equation
$\beta d^2 m/dx^2 = - dU/dm$ for a classical particle of mass
$\beta$ moving
in the potential $U(m) = - \alpha \sqrt{1 - m^2}$,
which can be solved by quadratures.
One thus obtains a one-parameter family of periodic profiles
$\zeta_S(x)$ which
are most conveniently parameterized by the maximum slope
$S \equiv \max_x \zeta_x$, and
which have been described
previously in the context of a different surface evolution equation
\cite{Srolovitz88}. The amplitude $A(S)$ 
\begin{figure}[t!]
\centerline{
 \epsfxsize=6cm  \epsfysize=4cm
  \epsfbox{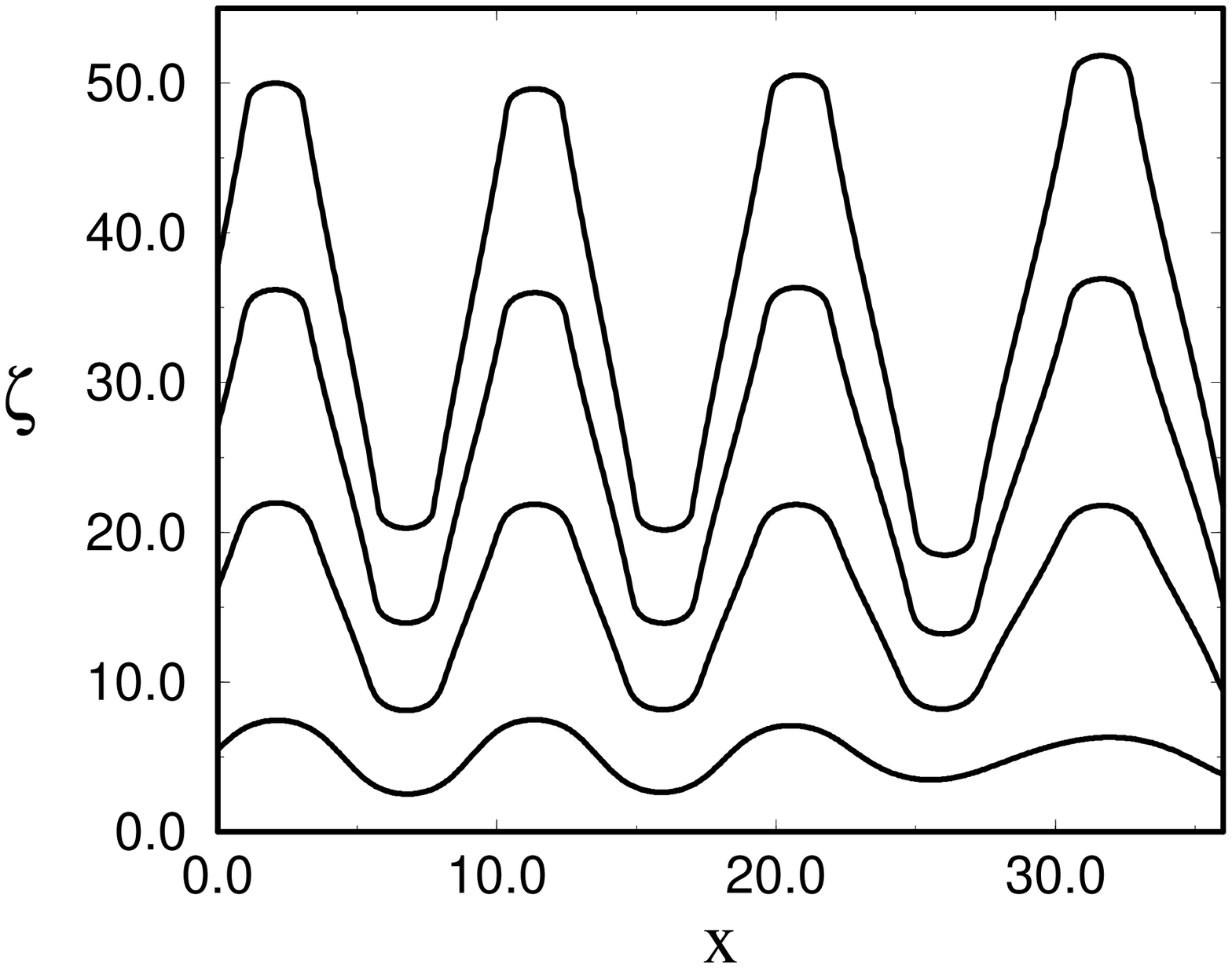}
}
\centerline{
 \epsfxsize=6cm  \epsfysize=4cm
  \epsfbox{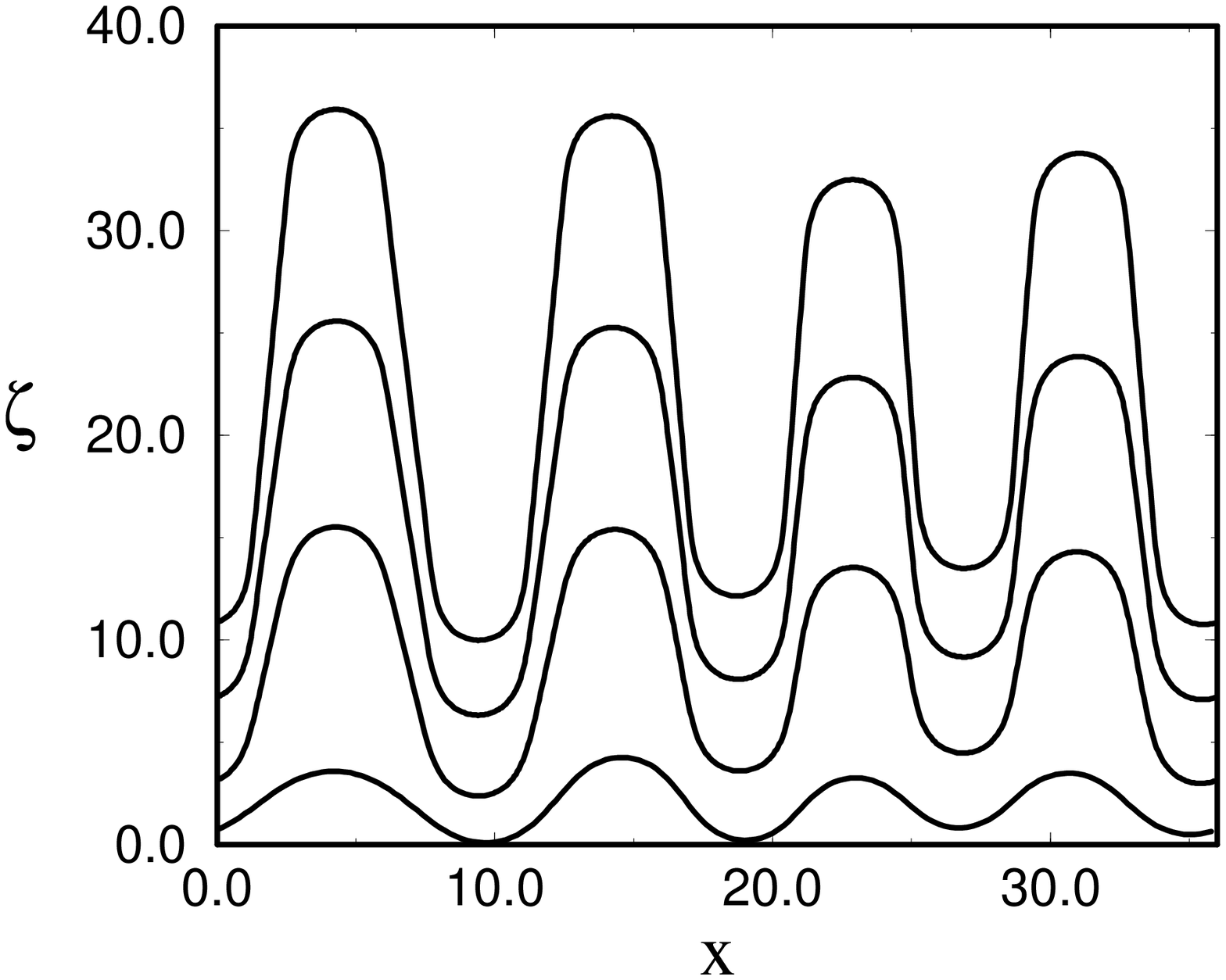}
}
\caption{
The evolution of the step profile starting from a flat
initial condition with small random
fluctuations. The upper figure shows the case of attachment-detachment
kinetics (Eq.(\ref{st_eq})) at times $t=36,64,110,183$,
the lower figure the case of edge diffusion
(Eq.(\ref{gen_eq}) with
$n=1/2$) at times $t=20,60,112,200$. Subsequent profiles have been shifted
in the $\zeta$-direction. In all figures
spatial variables have been scaled by $\lambda_c/2 \pi =
\sqrt{\beta / \alpha }$
and time by $\beta /\alpha^{2}$. }
\label{rand_prof}
\end{figure}
\noindent
is an increasing
function of $S$, while the wavelength $\Lambda(S)$ decreases
with increasing $S$, starting out at $\Lambda(0) = \lambda_c$.
For $S \to \infty$ finite limiting values $A(\infty) =
\sqrt{8 \beta/\alpha}$, $\Lambda(\infty) =
\sqrt{2 \pi \beta/\alpha} \; \Gamma(3/4)/\Gamma(5/4) \approx
0.5393527..\lambda_c$ are approached.

The {\it separable} solution of interest reads
\cite{Pierre-Louis98b,Krug97}
\begin{equation}
\label{wed}
\zeta(x,t) = 2 \sqrt{\alpha t}\  {\mathrm{erf}}^{-1} \left(
1 - 4 \vert x \vert/\lambda_s \right), 
\end{equation}
$-\lambda_s/2 < x < \lambda_s/2$,
where ${\mathrm{erf}}(z) = (2/\sqrt{\pi}) \int_0^z dy
\; e^{-y^2}$, and the wavelength $\lambda_s$ is arbitrary.
Equation (\ref{wed}) solves (\ref{st_eq}) exactly in the limit
$t \to \infty$, when the second term on the right hand side
becomes negligible compared to the first, and the evolution
equation reduces to $\zeta_t = -(\alpha/\zeta_x)_x$.
The solution (\ref{wed}) is singular near the maxima
and minima, where it diverges as $\zeta \sim \pm \sqrt{\ln(1/
\vert x - x_0 \vert)}$, $x_0 = 0, \pm \lambda_s/2$.

In Figure \ref{rand_prof} we show results of a numerical solution
of (\ref{st_eq}), starting
from a small amplitude random initial condition.
To secure good numerical stability we used a fully implicit,
backwards Euler algorithm for integration.
The algorithm was implemented on an adaptive grid
in order to obtain  sufficient lateral resolution at the singular
points.

\begin{figure}
\centerline{
 \epsfxsize=6cm  \epsfysize=5cm
  \epsfbox{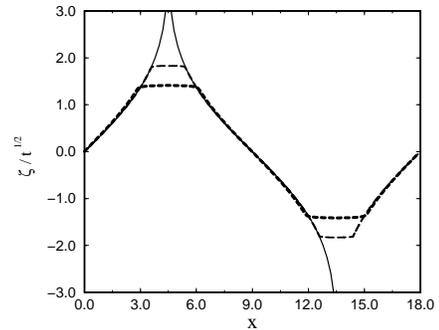}
}
\caption{The asymptotic form of the scaled profile
$\zeta/\sqrt{t}$ for Eq.(\ref{st_eq}) (long
dashes) and
Eq.(\ref{gen_eq}) with $n=1/2$ (short dashes). 
Full line is the separable solution
(\ref{wed}) with $\lambda_s$ equal to the total meander
wavelength $\lambda = 18 \sqrt{\beta/\alpha} \approx 2 \lambda_u$.}
\label{weds}
\end{figure}

\begin{figure}
\centerline{
 \epsfxsize=5cm  \epsfysize=4.5cm
  \epsfbox{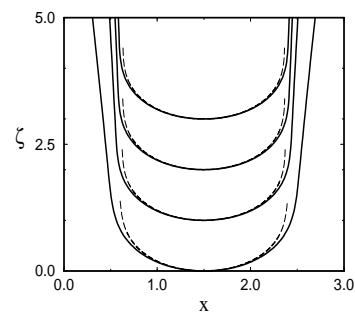}
}
\caption{The form of the caps for Eq.(\ref{st_eq})
at $t=500,1500,2500,3400$. The dashed lines are
the stationary solutions corresponding to the maximum slope in the profile.}
\label{caps}
\end{figure}

A regular meander pattern
of wavelength $\lambda_u$ develops, with an amplitude growing
indefinitely as $\sqrt{t}$. Closer inspection
reveals that the sides of the profile follow the
separable solution (Figure \ref{weds}),
while near the maxima and minima smooth {\it caps} appear
which approach pieces of the 
\noindent
stationary solutions
(Figure \ref{caps}).
This can be understood by noting that the mass current along
the sides of the profile vanishes as $1/\sqrt{t}$ according to
(\ref{wed}), and therefore the stationarity condition is
asymptotically satisfied; we have checked that the deviation from
the stationary profile which is discernible in Figure \ref{caps} vanishes
as $1/\sqrt{t}$. Since the slope of
(\ref{wed}) increases monotonically upon approaching an extremum while
it decreases for the stationary profiles, the matching of the two
solutions occurs near the point of maximum slope. For $t \to \infty$
the slope of the separable solution diverges, hence the cap
profile approaches the limiting stationary solution $\zeta_\infty(x)$,
and the length of the cap becomes $\Lambda(\infty)/2$. The rescaled step
profile $\zeta(x,t)/\sqrt{t}$ approaches an invariant shape in which
the cap appears as a flat facet. The wavelength $\lambda_s$ of the
separable solution depends on the cap length and on the total meander
wavelength $\lambda$, and is fixed by mass balance requirements
\cite{Gillet00};
for large total wavelength $\lambda_s \to \lambda$ (see
Figure \ref{weds}).
\begin{figure}
\centerline{
 \epsfxsize=6cm  \epsfysize=4cm
  \epsfbox{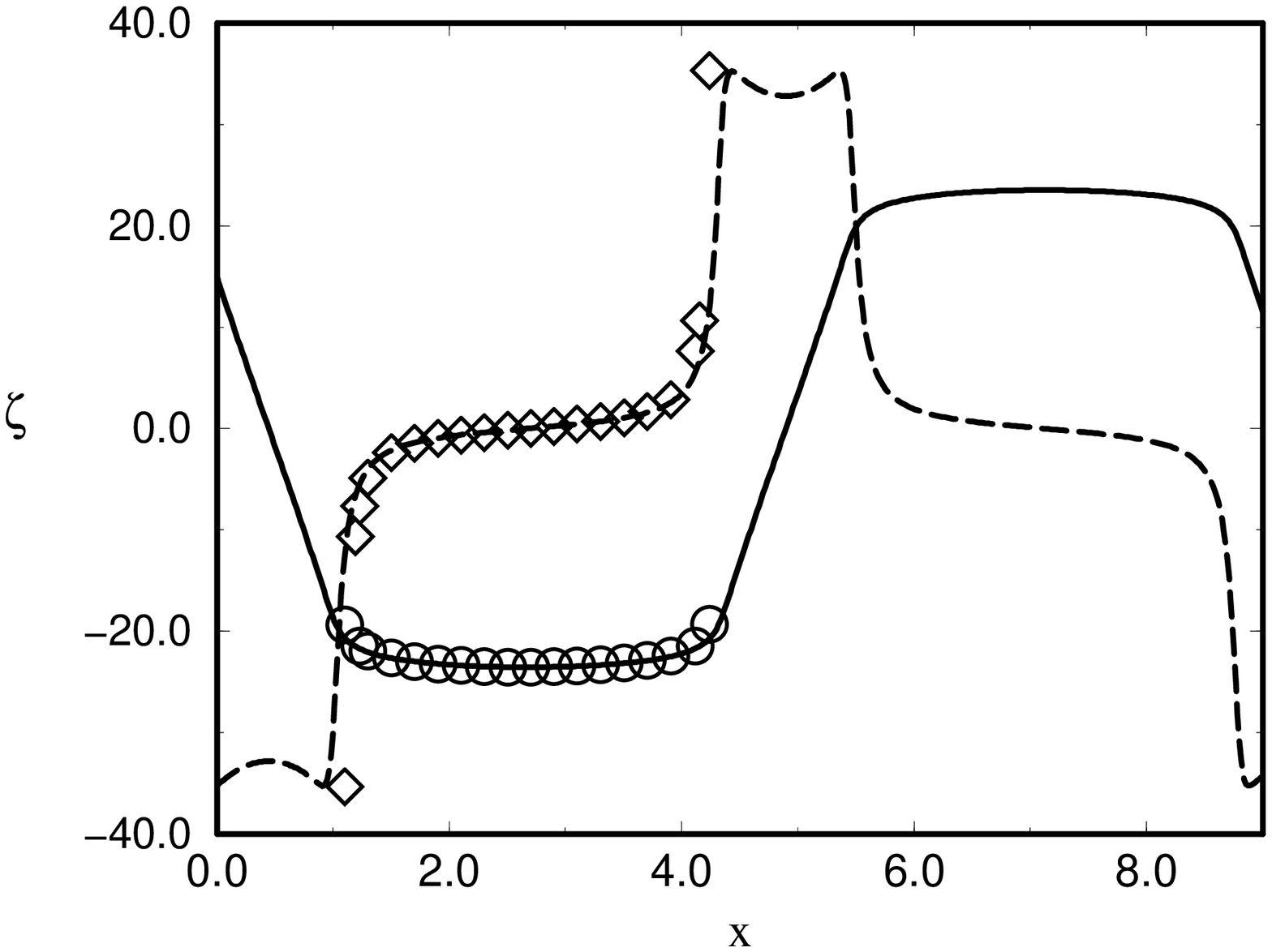}
}
\centerline{
 \epsfxsize=6cm  \epsfysize=4cm
  \epsfbox{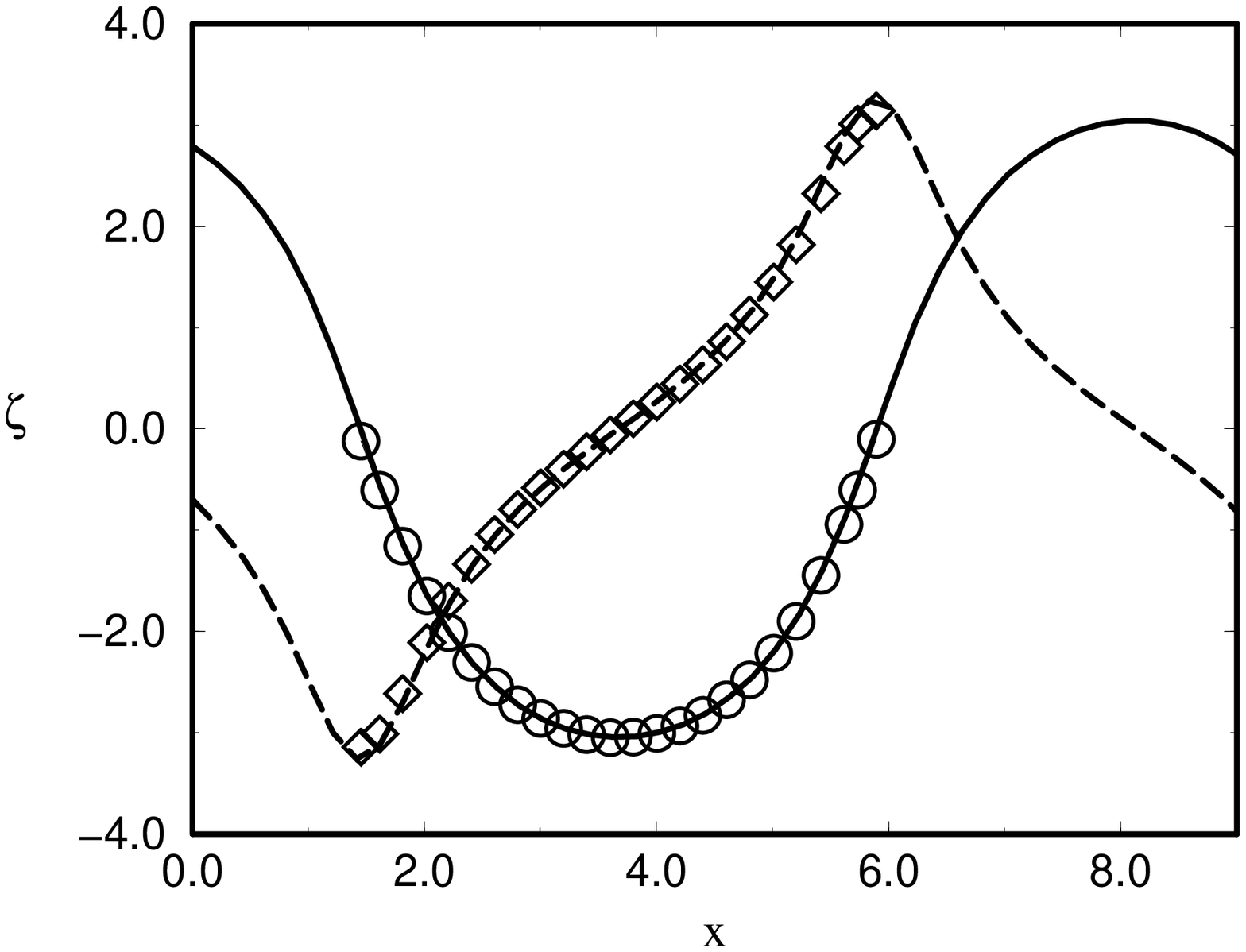}
}
\caption{Asymptotic form of the step profile for $n=1/2$ (upper) and
$n=0$ (lower). Full line is the profile and dashed line the
slope $\zeta_x$.
Circles represent the stationary solution and diamonds the
corresponding slope. The separable solution, still present in the
sloped regions of the profile for $n=1/2$, has vanished for
$n=0$.}
\label{as_prof}
\end{figure}

\section{Step edge diffusion and a generalized evolution equation}

On many fcc metal surfaces, diffusion along step edges is the fastest
kinetic process which therefore provides the dominant step smoothening
mechanism \cite{Jeong99}. To see how Eq.(\ref{st_eq}) has to be modified to
take
this effect into account, we note first that the second, relaxational
term on the right hand side can be rewritten in a geometrically
covariant form as $(\sigma \mu_s)_x$, where $\mu = \Omega \gamma \kappa$
is the step chemical potential \cite{Jeong99},
$\kappa = -  (1 + \zeta_x^2)^{-3/2} \zeta_{xx}$ is the step curvature,
$s = \int dx \; \sqrt{1 + \zeta_x^2}$
is the arclength along the step, and
\begin{equation}
\label{sigma}
\sigma = \frac{D \Omega c_{\rm eq}}{k_B T}
\frac{\ell}{\sqrt{1 + \zeta_x^2}}
\end{equation}
is a mobility.
The $\ell$-dependence in (\ref{sigma}) reflects
the assumed relaxation kinetics \cite{Pierre-Louis98b}, in which mass
exchange between different parts of the step occurs through detachment
followed by diffusion over the terrace, reflection at the descending
step, and re-attachment (case E of \cite{Pimpinelli93}).
The factor $1/\sqrt{1 + \zeta_x^2}$
has a simple geometric interpretation \cite{Krug99}:
For a deformed in-phase
step train the distance to the nearest step, measured along the step
normal, is $\ell/\sqrt{1 + \zeta_x^2}$ rather than $\ell$.

For relaxation
through step edge diffusion the mobility is clearly independent of
the step distance, and is given by \cite{Pimpinelli93}
$\tilde \sigma = D_{\rm e} \Omega c_{\rm e}/k_B T$, where $D_{\rm e}$
and $c_{\rm e}$
denote the edge diffusion coefficient and the
equilibrium concentration of edge
atoms, respectively. When edge diffusion dominates (i.e.
$\tilde \sigma \gg \sigma$), the appropriate nonlinear growth
equation should thus be given by (\ref{st_eq}) with the
second term replaced by $(\tilde \sigma \mu_s)_x =
([1 + \zeta_x^2]^{-1/2} \tilde \sigma \mu_x)_x$. This is confirmed
by the explicit derivation of Gillet {\it et al.} \cite{Gillet00}, who
also studied the crossover between attachment-detachment kinetics and
edge diffusion. Here our primary goal is to gain further insight into
the shape selection mechanism. This has
lead us to consider the generalized class of equations
\begin{equation}
\label{gen_eq}
\zeta_t = -\left\{ \frac{\alpha \zeta_x}{1+\zeta_x^2}  +
\frac{\beta}{(1+\zeta_x^2)^n}
\left[ \frac{\zeta_{xx}}{(1+\zeta_x^2)^{3/2}}
\right]_x \right\}_x,
\end{equation}
which reduces to (\ref{st_eq}) for $n=1$ and describes
relaxation through step edge diffusion when
$n=1/2$ and $\beta = \Omega^2 D_{\rm e} c_{\rm e} \gamma/k_B T$.
Below we discuss the properties of (\ref{gen_eq}) for general $n$,
keeping in mind that the cases
$n=1/2$ and $n=1$ are of immediate physical
relevance.

The separable solution (\ref{wed}) becomes exact in a limit
where the relaxation term in (\ref{st_eq}) can be neglected,
hence it remains a valid asymptotic solution also of (\ref{gen_eq})
for $n > -1/2$; for $n \leq -1/2$ the relaxation term can never
be ignored. The stationary solutions of (\ref{gen_eq}) can be
analyzed in terms of the same mechanical analogy described
above, the particle
potential being given by $U(m) = - \alpha (1 - m^2)^{3/2 - n}/(3-2n)$.
For $1/2 < n < 3/2$ the behavior is analogous to that
for $n=1$: The wavelength $\Lambda(S)$ is a decreasing function
of the maximal slope $S$, and wavelength and amplitude
reach finite values $A(\infty) = \sqrt{8 \beta/\alpha}
\sqrt{3 - 2 n}/(2n-1)$ and
\begin{equation}
\label{Lambda}
\Lambda(\infty) = \sqrt{2 \pi(3 - 2n)(\beta/\alpha)} \frac{
\Gamma[(2n+1)/4]}{\Gamma[(2n+3)/4]}
\end{equation}
for $S \to \infty$. Thus the
asymptotic step profiles look similar to those
generated by Eq.(\ref{st_eq}),
with the length of the cap decreasing with increasing
$n$. As $n \to 3/2$ the cap length, given by $\Lambda(\infty)/2$, vanishes.
For $n \geq 3/2$ we therefore expect true
spike singularities to develop at
the maxima and minima of the profile. Using that the
slope imposed by the separable solution (\ref{wed}) grows as
$S \sim \sqrt{t}$, we predict that the curvature at the extrema diverges
as $t^{(2n-3)/4}$.

A numerical solution for the case of edge diffusion ($n=1/2$) is shown
in the lower panel of Figure \ref{rand_prof}.
For $n=1/2$ the potential $U(m)$ is harmonic and hence the
wavelength $\Lambda(S) = \lambda_c$ independent of $S$. The amplitude
of the stationary profiles diverges as $A(S) \sim \ln S$, leading
to a corresponding increase of the cap height as $\ln t$. Since
this is still small compared to the overall profile amplitude,
the caps nevertheless appear as flat in the
rescaled shape $\zeta/\sqrt{t}$
(Figure \ref{weds}; a detailed view
of the cap is shown in Figure \ref{as_prof}).
This remains true in the entire interval
$-1/2 < n \leq 1/2$, where the cap length (\ref{Lambda}) remains
finite and the cap height grows as $t^{(1-2n)/4}$. However a qualitative
change in the profile evolution occurs at the value
$n_c \approx 0.2283$ where the asymptotic stationary
wavelength (\ref{Lambda})
becomes equal to the most unstable wavelength $\lambda_u$, which
sets the lateral length scale in the 
\begin{figure}
\centerline{
 \epsfxsize=7cm  \epsfysize=5cm
  \epsfbox{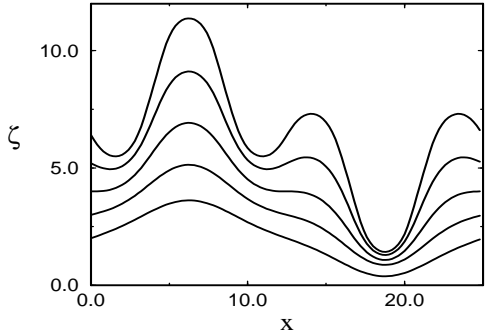}
}
\caption{Spontaneous creation of an extra meander period for
the case of step-edge diffusion ($n=1/2$).
The initial wavelength of the profile is $\lambda_i= 25
\sqrt{\beta/\alpha} >
3 \lambda_c$.}
\label{per_crea}
\end{figure}
\noindent
early stages of growth.
For $n < n_c$ we expect to see an intermediate {\it coarsening}
regime in which the lateral length scale increases from $\lambda_u$
to $\Lambda(\infty)$. Coarsening to arbitrarily large length scales,
similar to what is observed in related evolution equations for
one-dimensional unstable growth \cite{Politi00b}, sets in at $n = -1/2$,
where (\ref{Lambda}) diverges. Throughout the regime $n < n_c$ the
evolving profile is describable in terms of stationary solutions, and
the separable solution (\ref{wed}) no longer plays any role
(Figure \ref{as_prof}).

\section{Persistence of the initial wavelength}

Finally, we address the
recent experiments \cite{Maroutian99}
on surfaces vicinal to Cu(100), in which the meander
wavelength was measured as a function of temperature, and it was
concluded that the observed behavior is {\it inconsistent} with the
theoretical prediction for the linearly most unstable wavelength
$\lambda_u$. Maroutian {\it et al.} \cite{Maroutian99}
therefore proposed that the meander wavelength is set by the
{\it nucleation length} describing the distance between the one-dimensional
nuclei appearing on a flat step in the early stage of growth, which can
be considerably larger than $\lambda_u$.

A necessary consistency requirement for this scenario is that an initially
imposed meander wavelength $\lambda_i > \lambda_u$ persists during the
nonlinear evolution. We have therefore numerically integrated
Eqs.(\ref{st_eq}, \ref{gen_eq}) starting from a sinusoidal initial condition
with varying wavelength $\lambda_i$. We do find that a range of wavelengths
can be preserved during growth. This is reasonable in view of the analysis
presented above, which shows that asymptotic profiles, composed of
the separable solution (\ref{wed}) and a stationary cap, can in principle
be constructed for arbitrary wavelength
(see e.g. Figure \ref{weds}). However, when $\lambda_i$ exceeds
$\lambda_c$ by more than a factor of 3, so that an additional meander
fits between the maxima and minima of the profile, the wavelength spontaneously
decreases to a value near $\lambda_u$ (Figure \ref{per_crea}).
This result contradicts the assumption of \cite{Maroutian99} that
initial wavelengths much larger than $\lambda_u$ persist, but
it should
not be overemphasized: Clearly processes which involve
a change in the collective meander wavelength may not
be accurately described in a model which assumes in-phase meandering
from the outset.

\section{Outlook}

In conclusion, we have described an unusual shape selection scenario
for a class of physically motivated growth equations. A number of
issues remain to be clarified. Mathematically, the behavior in
the region where separable and stationary solutions match needs
further investigation; our numerical work indicates the appearance
of singularities in higher derivatives of $\zeta$. Also the
dynamics in the singular
regime $n \geq 3/2$ and in the coarsening regime $n < n_c$
of Eq.(\ref{gen_eq}) deserves attention. Physically, it is
imperative that the predictions of the one-dimensional equations for
the in-phase step meander be confirmed by more complete descriptions
of the growing surface, as provided by two-dimensional continuum
equations and Monte Carlo models \cite{Rost96}, in order to
assess their ultimate relevance for the experimentally observed
morphologies.

\section*{Acknowledgements}

We are much indebted to Jens Eggers for help with the numerical
algorithm and useful discussions. Olivier Pierre-Louis and Chaouqi Misbah
kindly supplied us with a copy of \cite{Gillet00} prior
to publication. Support by DFG/SFB 237
is gratefully acknowledged.

\end{multicols}


\begin{references}
\bibitem{Note:Gillet1} An independent numerical investigation
was carried out in Ref.\cite{Gillet00}. Their conclusions agree
with ours.
\end{references}
\end{document}